\documentclass[twocolumn,prb]{revtex4}

\def\beq{\begin{equation}}
\def\eeq{\end{equation}}
\def\Li{LiCu$_2$O$_2$}
\def\Na{NaCu$_2$O$_2$}
\def\LiV{LiVCuO$_4$}

\usepackage{graphicx}
\usepackage{epsfig}

\begin{document}

\title{Spin polarization of the magnetic spiral in \Na \, as seen by NMR}
\author{A.A. Gippius}
\affiliation{Moscow State University, 119899, Moscow, Russia}
\author{A.S. Moskvin}
\affiliation{Ural State University, 620083, Ekaterinburg, Russia}
\author{S.-L. Drechsler}
\affiliation{Leibniz Institut f\"ur Festk\"orper- und Werkstofforschung
Dresden, D-01171, Dresden, Germany}

\date{\today}
\begin{abstract}

The incommensurate (IC) spin ordering in  quasi-1D edge-shared cuprate  \Na \, has been studied by $^{23}$Na nuclear magnetic resonance spectroscopy in an external magnetic field near 6 Tesla applied along the main crystallographic axes. The NMR lineshape evolution above and below  $T_N$$\approx$12 K yields  a clear signature of an IC  static modulation of the local magnetic field consistent with a Cu$^{2+}$ spin spiral polarized in the $bc$-plane rather than in the $ab$-plane as reported from earlier neutron diffraction data. 

\end{abstract}


\maketitle

The magnetic phase transitions observed at low temperature in several edge-shared chain cuprates (e.g., \Li , \LiV , \Na ) are considered as evidence for an incommensurate (IC) helicoidal order with  propagation along the chain direction.\cite{Drechsler}  This picture based on strong in-chain frustration is supported by NMR and neutron diffraction  measurements.\cite{Gippius,Masuda,Capogna,Berthier} However, many significant details of this spin-ordered state  have not been settled so far. In particular, the orientation of the spin rotation (within the probably simplified picture of a planar spiral) is under hot debate due to the recent observation of a multiferroic behaviour induced by the spin ordering in \LiV \, and \Li.\cite{Naito,Naito1,Schrettle,Cheong,Seki} In fact, the appearance of a spiral itself and  its orientation are of crucial importance for all proposed mechanisms and phenomenological approaches to multiferroicity.\cite{Mostovoy} However, multiferroic behavior in  \Na \, isomorphic to \Li \, has not been reported. Hence, the spiral order seems to be not directly related with multiferroicity. 
For both \Li \, and \Na \, an
$ab$ spin polarization has been deduced  from earlier neutron
 diffraction measurements.\cite{Masuda,Capogna} That was also supported for \Li \, by later
 ESR data.\cite{Mihaly}
However, the observation of spontaneous ferroelectric polarization $\bf P$$\parallel$$c$-axis below $T_N$ by Park {\it et al.}\cite{Cheong} raised  doubts of the $ab$ spin polarization in \Li \, in favour of a $bc$ spin polarization, which was partially supported by very recent neutron diffraction measurements by Seki {\it et al.}\cite{Seki} 
To find support for the $bc$ polarization   these authors have pointed to the paper by Capogna {\it et al.}\cite{Capogna} on the isomorphic \Na \, where  contradictory orientations have been reported.\cite{Keimer} 
Thus we arrive at the puzzling situation that we have seemingly no reliable data regarding the spin polarization in two isomorphic IC chain cuprates \Li \, and \Na. 
 The present-day  \Li \, samples still exhibit  significant nonstoichiometry with  both nonmagnetic Li impurities
in the CuO$_2$ chains, and magnetic Cu$^{2+}$ impurities positioned in between chains.\cite{Masuda} These impurities have not  been considered in previous papers. However, the recently  observed multiferroic behavior in \Li \, can be consistently 
explained, if  the exchange-induced electric polarization on the
out-of-chain Cu$^{2+}$ centers substituting for Cu$^{+}$-ions are taken into account\cite{multi-LiCuO} (see also Ref.\,\onlinecite{LiV} on \LiV ). Interestingly, regular  spiral chains spin-polarized in $ab$-plane induce on these Cu$^{2+}$ centers a spin polarization along $c$-axis. This can explain some seeming inconsistencies found recently in neutron diffraction data \cite{Seki} but without any all-out negation of an $ab$-plane spiral. 
Due to a larger ionic radius of Na$^{+}$ (0.97 \AA \, versus 0.68 \AA \, of Li$^{+}$) substitutional disorder is {\it a priori} unlikely in \Na \, and we deal here with a higher degree of in-chain crystallographic order and hence increasing one-dimensionality of magnetic properties. In contrast with \Li, the \Na \, single crystals exhibit  no twinning and  no deviation from the ideal stoichiometry as confirmed by X-ray and TGA-analysis. The $^{63,65}$Cu NQR lines in \Na \, in the paramagnetic state are a
factor of 3 more narrow than those in \Li \, reflecting the higher
degree of crystallographic order.\cite{JMMM} 
 Likely, \Na \, samples are more relevant to compare the data provided by different techniques. 
The $^{63,65}$Cu NMR (Ref.\,\onlinecite{Berthier}) and our preliminary $^{23}$Na NMR
data\cite{JMMM} have confirmed the IC ordering in \Na . However, both groups did not
cast doubt on the $ab$-plane spin spiral polarization reported earlier in Ref. \onlinecite{Capogna} on the basis of neutron diffraction. Below in the paper we report comprehensive data of $^{23}$Na NMR measurements in   \Na \, for different field orientations and provide strong arguments for the $bc$- rather than the $ab$-plane spin polarization in  contrast with that neutron data.\cite{Capogna,Keimer}

The single crystalline samples of orthorhombic   \Na \,     used in our experiments were grown as described in Ref.\,\onlinecite{Masuda}.
\begin{figure}[t]
\includegraphics[width=7.0cm,angle=0]{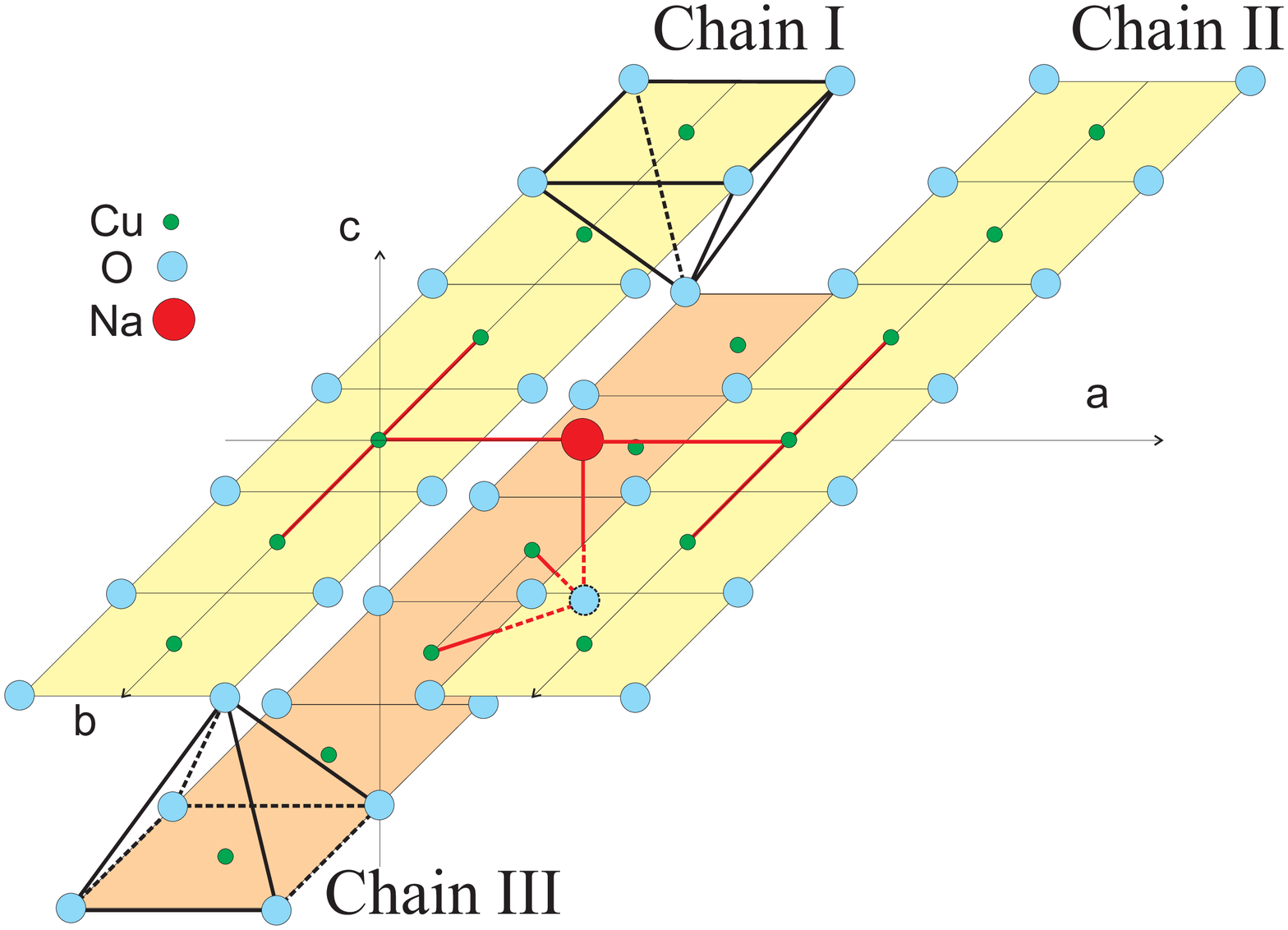}
\caption{(Color online) Schematic view of the active
triple Cu$^{2+}$O$_2$ chain structure of \Na. 
The nonmagnetic Cu$^{+}$ ions are omitted.
The  
hyperfine coupling geometry is shown by bold lines.} \label{fig1}
\end{figure}
The unit cell\cite{Maljuk} contains four magnetic Cu ions belonging to two pairs of CuO$_2$ chains running along $b$-axis and interconnected by Cu$^{1+}$ in O-Cu-O dumbbells (Fig.\,\ref{fig1}). 
 NaCu$_2$O$_2$  is a magnetic insulator with  a magnetic phase transition to a spiral state below  $T_N$=12-13 K.\cite{Capogna,Drechsler1} 
 The first experimental evidence of magnetic IC order in isostructural \Li \, and \Na \,was obtained independently by Gippius {\it et al.}\cite{Gippius}  and Masuda {\it et al.}\cite{Masuda} for \Li\, from $^{6,7}$Li NMR  and neutron diffraction measurements, respectively, and by Capogna {\it et al.}\cite{Capogna} and Horvati\'{c} {\it et al.}\cite{Berthier} for \Na \, from the neutron diffraction measurements and $^{63,65}$Cu NMR, respectively.
The reported  fit of the neutron data\cite{Masuda,Capogna} means that all spins are confined to the  $ab$  plane and form a planar spin helix $\bf S_i=S(\cos\theta_i,\sin\theta_i,0)$, where $\theta_i={\bf q} \cdot {\bf r}_i+\alpha$,  $\alpha$ is a phase shift, and ${\bf q} $ is the propagation vector.
Actually, both spin anisotropy and external magnetic field  may distort a classical spin helix. Anisotropy in the spin plane perturbs the spin helix adding higher-order even harmonics.\cite{Nagamiya} An external magnetic field applied perpendicular to the spin helix plane (transverse field) preserves the in-plane spiral order and induces  spin canting towards the field direction  producing  an umbrella spin structure of the form
$\bf S_i=S(\cos\gamma\cos\theta_i,\cos\gamma\sin\theta_i,\sin\gamma )$ with $\sin\gamma=H/H_s$ which remains up to the saturation field $H_s$. Its effect on the NMR line shape reduces to a decrease in splitting $\propto \cos\gamma$, and to a rigid shift $\propto \sin\gamma$.
An external magnetic field applied in the spin plane (longitudinal field) perturbs the spin helix adding odd higher-order harmonics.\cite{Nagamiya} For the  easy axis and the external field both directed along the  $a$-axis  we arrive at the transformation 
\begin{equation}
\theta_i={\bf q}_i\cdot {\bf r}_i -\theta_H \sin({\bf q}_i\cdot {\bf r}_i)-\theta_{an} \sin(2{\bf q}_i\cdot {\bf r}_i)\, ,
\label{bunch}
\end{equation}
where the deviation angle $\theta_{H} $ depends linearly  on the external field. 

 The hyperfine (HF) field induced by a classical planar spin helix  on a nucleus positioned at site ${\bf R}$ near a CuO$_2$  chain is directly related to the local spin polarization on the neighboring sites ${\bf S}({\bf 
R}+{\bf r})$: ${\bf h}({\bf 
R}) =\sum_{{\bf r}}{\hat A}({\bf r}){\bf S}({\bf R}+{\bf r})$, where ${\hat 
A}({\bf r})$ is the anisotropic HF tensor taking into account the magnetic dipole and  the supertransferred Cu-O-Na HF interactions. The local field on a Na nuclei is induced  by a superposition of at least three neighboring spin spirals (I,II,III in Fig.\,1). 
 The local field on the $^{23}$Na nuclei which is induced by the isotropic supertransferred  HF interaction from the antiferromagnetically coupled chains is strongly canceled. This  makes the anisotropic interaction a main contributor. Moreover, symmetry considerations point to $A_{ab}$ as being the only nonzero component of the net HF  tensor for coupling to chains I and II.  
Generally speaking, the HF field on an out-of-chain nucleus  can be written as follows:
\begin{equation}		
	h_{x,y,z}=A_{x,y,z}(q)\cos(qy+\alpha_{x,y,z})\, ,
\end{equation}
with the effective HF coupling parameters $A_{x,y,z}$ and the HF phase shifts $\alpha_{x,y,z}$ which may differ from the spin spiral phase shift $\alpha$. 
In a  continuum approximation the resultant NMR line shape associated with a single nuclear $\Delta m_I=\pm 1$ transition   can be calculated straightforwardly by a simple summation (integration):
\begin{equation}
	F({\bf H})\propto\int_0^{2\pi}\exp(-(|{\bf H}+{\bf h}(\phi)|-H_{L})^2/2\delta ^2)d\phi \, ,
\end{equation}
 where ${\bf H}$ and $H_{L}$ are the external and the resonance Larmor fields, respectively, $\phi =qy$, and $\delta$ denotes the homogeneous line width.
A symmetrically bunched umbrella-like spin spiral for the transversal field geometry provides a symmetric line shape of the NMR response, while an asymmetrically bunched  spin spiral for the longitudinal, or in-plane  field geometry  yields an asymmetric NMR line shape.\cite{Blinc} This simple relation can be used for a fast assignment of the spin spiral polarization. It should be noted that the character of the NMR lineshape asymmetry depends both on the sign of the HF coupling constant and the phase shift. 
\begin{figure}
\includegraphics[width=7.5cm]{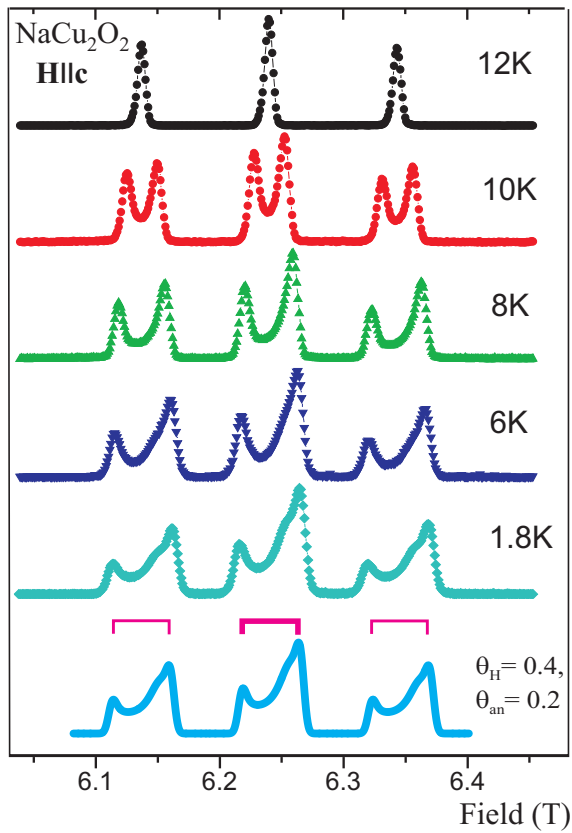}
\caption{(Color online) The $^{23}$Na NMR spectrum   for {\bf H}$\|${\bf c} with resonance frequency of 70.0 MHz. Bottom: Theoretical simulation with a simple asymmetrically bunched plane spiral ($\Delta_M^c=0.06, \nu _Q^c=0.103$\,T). Rectangular brackets indicate here and below in Figs.\,3 and 4 the magnetic splitting of the central (thick) and satellite transition lines.
} \label{fig2}
\end{figure}

 We have performed $^{23}$Na NMR measurements on a 
NaCu$_2$O$_2$ single crystal in the paramagnetic and in the ordered 
phase by sweeping an external field{\bf H} at  fixed frequency of 70.0 MHz. The external field was oriented along the main crystal axes {\bf H}$\|${\bf a},{\bf b},{\bf c}. The signal was obtained by integrating the spin-echo envelope.  The experimental spectra are presented in Figs.\,2-4. In order to discuss them we start  with the case of a {\bf H}$\|${\bf c} geometry. At a first glance this implies a relatively simple symmetric picture of  $^{23}$Na NMR signal induced by  $ab$-plane polarized spin helix as reported in Ref.\,\onlinecite{Capogna}. Indeed, in this case an  external magnetic field applied perpendicular to the spin helix plane should preserve the in-plane spiral order and should induce a spin canting towards the field direction resulting in  a rigid shift of NMR frequency. However, our experimental results point to a completely different picture as we will explain below. 
At $T$$>$$T_N$ a  first order quadrupole perturbed NMR spectrum 
typical for spin $I$=3/2 was observed (Fig.\,\ref{fig2}). It contains 3 lines, nearly equally spaced by a quadrupolar coupling to the local electric field gradient. The central line and the two satellites show an intensity ratio close to the theoretically expected one: 3:4:3.
The quadrupole splitting  $\nu _Q^c=0.103$\,T does not reveal a noticeable temperature dependence. 
	But for $T$$ <$12 K a dramatic change of the $^{23}$Na NMR spectrum is 
observed with a continuous and identical splitting of  the quadrupole triplet components.
 This is a text-book signature of an infinite number of magnetically non-equivalent $^{23}$Na sites typical for an IC static modulation of the local magnetic fields.\cite{Blinc} Indeed, the magnetic component of the IC lineshape is dominantly given by the central transition. If the same lineshape is observed for all the lines, the quadrupolar coupling is not modulated, and we conclude that the origin of the modulation is only magnetic. Obviously, such a   static IC modulation of the local magnetic field is caused  by the helical spin structure of the Cu magnetic moments similarly to the structure observed in LiCu$_2$O$_2$.\cite{Gippius} At variance with the $^{7}$Li NMR in lithium cuprate, the $^{23}$Na NMR spectrum in \Na \, shows  a nice picture of  practically identical lineshapes of the satellite transitions, which  are here 
clearly observable due to a comparable magnitude of  quadrupole and magnetic splittings. 
However, the $^{23}$Na NMR lineshape is strongly asymmetric, especially at low temperatures (Fig.\,2).
 It  cannot be  explained in the framework of an $ab$-plane spin polarization. Instead, it points to the NMR response of an $ac$- or $bc$-plane  polarized spin spiral. In fact,    the $^{23}$Na NMR lineshape at T=1.8 K can be successfully simulated assuming such a situation ($\theta _{an}=0.2,\theta _{H}=0.4$) with  a magnitude of a magnetic splitting $\Delta_M^c=2|A_{z}|=0.06$\,T (Fig.\,\ref{fig2}). 
\begin{figure}
\includegraphics[width=7.0cm]{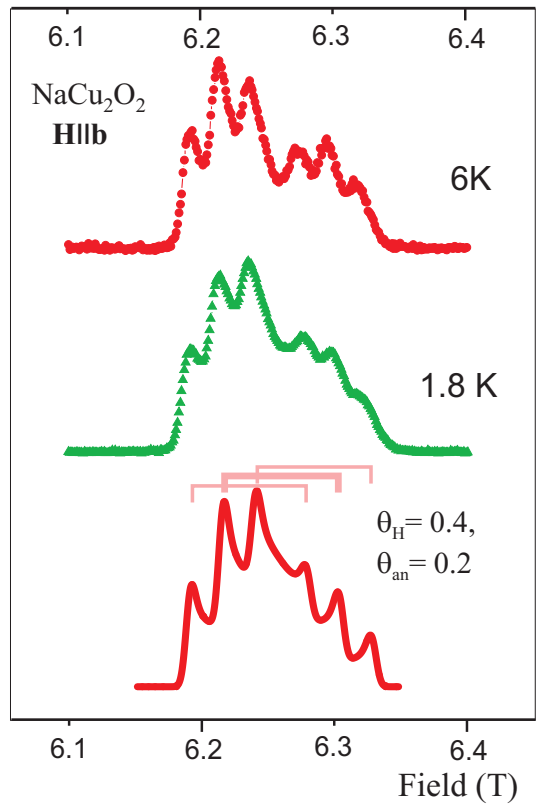}
\caption{(Color online) The $^{23}$Na NMR spectrum  below 
 $T_N$ for {\bf H}$\|${\bf b} with resonance frequency of 70.0 MHz. Bottom: theoretical simulation  with a simple asymmetrically bunched  plane spiral ($\Delta_M^b=0.09, \nu _Q^b=0.022$\,T). } \label{fig3}
\end{figure}
\begin{figure}
\includegraphics[width=7.5cm]{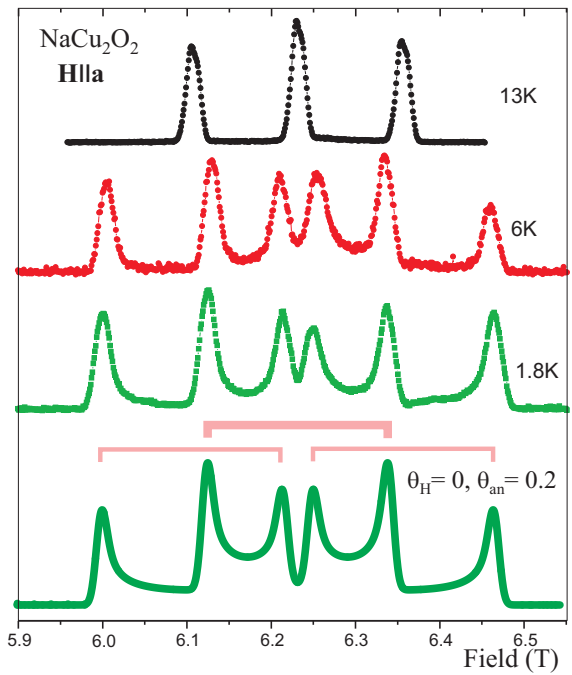}
\caption{(Color online) The $^{23}$Na NMR spectrum  below 
 $T_N$ for {\bf H}$\|${\bf a} with resonance frequency of 70.0 MHz. Bottom: theoretical simulation with a simple symmetrically bunched  $bc$-plane spiral ($\Delta_M^a=0.22, \nu _Q^a=0.125$\,T). } \label{fig4}
\end{figure}
Of course, such a proposal has to be tested with an external field applied along the $a$- and $b$-axes, respectively.

 The {\bf H}$\|${\bf b}-axis NMR spectrum (Fig.\,\ref{fig3}) shows a noticeable asymmetry of the NMR lineshape  which clearly implies a non-$ac$-plane spin  polarization. 
The quadrupole splitting  $\nu _Q^b=0.022$\,T is  small as compared with  $\nu _Q^{c}$, however, the magnetic splitting saturates at an intermediate value of 0.09 T. A clear visual detection of magnetic splitting is hindered by the  small quadrupole splitting. We found that the $^{23}$Na NMR lineshape can be    simulated  within  a "single bunched spiral model" with parameters: $\Delta ^b_M=2|A_{y}|=0.09$\,T, $\nu^b_Q = 0.022$\,T, $\theta _{H}=0.4$, $\theta _{an}=0.2$.

Thus, after excluding the possibility of $ab$ and $ac$ polarization planes
based on the discusion of the measured NMR spectra in {\bf H}$\|${\bf c} and {\bf H}$\|${\bf b} geometries, only the $bc$-plane remains as a possible  spin polarization plane.
Indeed,  for an external field applied along the $a$-axis {\bf H}$\|${\bf a} we arrive at a highly symmetric $^{23}$Na NMR lineshape (Fig.\,\ref{fig4}) which can be well described in terms of a simple planar spiral model with $\Delta ^a_M=2|A_{x}|=0.22$\,T, $\nu^a_Q = 0.125$\,T, $\theta _{H}=0$, $\theta _{an}=0.2$. 
 The quadrupole splitting\cite{Laplace}   is slightly larger than $\nu _Q^c$, however, the magnetic splitting saturates at a considerably larger value of 0.22\,T. Therefore, in an {\bf H}$\|${\bf a} orientation the magnetic splitting overlaps the quadrupole splitting in contrast to the {\bf H}$\|${\bf c} geometry (Fig.\,\ref{fig2}). 
 Finally, we see that not an $ab$- but a $bc$-plane polarized spin helix is  robust with respect to the application of a rather strong external magnetic field of about 6\,T. The magnetization measurements for another sample of the same batch  in an external field along $a,b,c$ axes shows no signatures of spin-orientational transitions for the fields up to 7\,T (Fig.\,2 in Ref.\,\onlinecite{Drechsler1}). Obviously, the $bc$-plane is an  easy spin plane in \Na. Of course, a spin-flop transition  is expected to occur too but for higher fields. 
Starting with the $bc$-plane polarized spin spiral we have estimated the magnetodipole contribution to the maximal magnetic splitting of the $^{23}$Na NMR line to be $\Delta _M^a=0.109$\,T/$\mu _B$, $\Delta _M^b=0.050$\,T/$\mu _B$, $\Delta _M^c=0.027$\,T/$\mu _B$ given the   pitch angle 81.7$^{\circ}$.  The magnetodipole  mechanism predicts qualitatively correctly the anisotropy of the magnetic splittings, however, its  contribution explains only  a fourth of the net effect given the  magnetic moment $\sim$0.6 $\mu_B$ per Cu$^{2+}$ ion.\cite{Capogna} In other words,  our experimental data point to a significant effect of anisotropic Cu-O-$^{23}$Na supertransferred HF coupling which seems to be more pronounced as compared with the similar  Cu-O-$^{7}$Li bonds in Li$_2$CuO$_2$.\cite{Giri}

In conclusion, 
the $^{23}$Na NMR lineshape in
NaCu$_2$O$_2$ shows  clear signatures of an IC static spin structure
 consistent with a spiral modulation of the
Cu  magnetic moments   polarized in the $bc$-plane in contrast with the $ab$-plane polarization reported earlier. It is the first experimental indication for a   polarization in an edge-shared cuprate with spins lying in a plane perpendicular to the plane of the basic CuO$_4$ plaquette.
We have found the values of magnetic $\Delta _M^{a,b,c}$ and quadrupole $\nu _Q^{a,b,c}$  splittings which provide a starting database for a further detailed study of magnetic and electric HF interactions in this cuprate. 
 Our results obtained for a clean system should be of considerable interest
 both for the magnetic anisotropy in general and  the symmetry aspects of possible multiferroicity\cite{Mostovoy,mechanism}  in chain cuprates.

 We thank M. Baenitz, A. Zheludev,  B. Keimer and E. Morozova for  useful  discussions, and  K. Okhotnikov for an assistance with the experiment. Partial support by  RFBR grants Nos. 06-02-17242, 07-02-96047, and 08-02-00633 (ASM),  Russian Science Support Foundation
 (AAG), and the DFG (S.L.D) is  acknowledged.


\begin{thebibliography}{99}
\bibitem{Laplace}
Quadrupole splittings $\Delta_Q^{a,b,c}$ derived from NMR spectrum   agree with the Laplace equation: $V_{xx}+V_{yy}+V_{zz}=0$, where ${\hat V}$ is an electric field gradient.
\end{thebibliography}
\end{document}